\documentclass[aps,a4paper,10pt,twocolumn,nofootinbib,floatfix]{revtex4} 
\usepackage{amsmath}
\usepackage{amsfonts}
\usepackage{amssymb}
\usepackage{mathrsfs}
\usepackage[mathscr]{euscript}
\usepackage[dvips]{graphicx}
\usepackage{fancyhdr}
\usepackage{makeidx}
\usepackage{colordvi}

\def\overstrike#1#2{{\setbox0\hbox{$#2$}\hbox to \wd0{\hss
    $#1$\hss}\kern-\wd0\box0}}

\begin{document}

\title{Transverse limits on the uni-directional pulse propagation approximation}

\author{P. Kinsler}
\email{Dr.Paul.Kinsler@physics.org}

\affiliation{
  Blackett Laboratory, Imperial College London,
  Prince Consort Road,
  London SW7 2AZ,
  United Kingdom.}

\begin{abstract}

I calculate the limitations on the widely-used forward-only
 (uni-directional) propagation assumption
 by considering the effects of transverse effects (e.g. diffraction).
The starting point is the scalar second order wave equation, 
 and simple predictions are made which aim to 
 clarify the forward-backward coupling limits on diffraction strength.
The result is unsurprising, 
 being based on the ratio of transverse and total wave vectors, 
 but the intent is to present a derivation 
 directly comparable to a recently published \emph{nonlinearity} constrained 
 limits on the uni-directional approximation \cite{Kinsler-2007josab}.

\end{abstract}

\lhead{\includegraphics[height=5mm,angle=0]{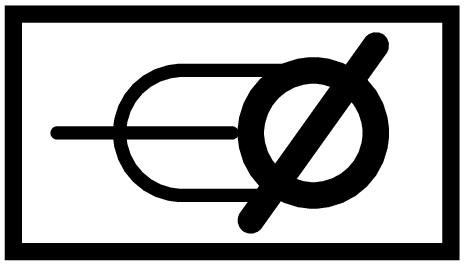}~~FBDIFF}
\chead{Transverse limits to the uni-directional ...}
\rhead{
Dr.Paul.Kinsler@physics.org\\
http://www.kinsler.org/physics/
}

\date{\today}
\maketitle
\thispagestyle{fancy}

\section{Introduction}
\label{S-Introduction}

Most approaches to optical pulse propagation rely on an 
 approximation where the fields
 only propagate forwards.
Even the recently derived extensions of 
 typical propagation methods used in nonlinear optics
 (e.g. 
 \cite{Brabec-K-1997prl,Kinsler-N-2003pra})
 assume a complete decoupling between 
 oppositely propagating fields
 to optimize the calculation.
Moreover,
 those based directly on Maxwell's equations
 (e.g. \cite{Kolesik-MM-2002prl,Tyrrell-KN-2005jmo,Kinsler-RN-2005pra})
 or the second order wave equation 
 (e.g. \cite{Blow-W-1989jqe,
             Ferrando-ZCBM-2005pre,Genty-KKD-2007oe}),
 are often simplified to work in the forward-only limit, 
 where backward propagating fields are set to zero.
This is despite 
 directional decompositions of Maxwell's equations 
 (e.g. \cite{Kinsler-RN-2005pra,Kinsler-2006arXiv-fleck})
 indicating that \emph{any} effect not allowed for by that 
 decomposition couples the forward and backward waves together --
 and even creates a backward field if one is not present.
Usually
 we assume that a forward wave will not 
 generate a significant backward wave via the diffraction
 because we are in the paraxial limit, 
 and even then any generated backward component is very poorly phase 
 matched\footnote{
  If the forward field has a wave vector $k_0$ 
   evolving as $\exp(+\imath k_0 z)$, 
   the generated backward component will evolve as $\exp(-\imath k_0 z)$.
  This gives a very rapid relative oscillation $\exp(-2\imath k_0 z)$, 
   which will quickly average to zero.}.

I compare predictions for propagation wave vector
 from uni- and bi-directional theories.
Such a comparison has been done for the more important case
 of nonlinearity-induced forward-backward coupling
 \cite{Kinsler-2007josab}, 
 and used analytical expressions for carrier shocking 
 \cite{Rosen-1965pr,Kinsler-RTN-2007pre}
 in concert with simulations 
 to examine the effects of a uni-directional approximation.
Here my intention is simply to provide a complementary calculation 
 to clarify the effects of forward-backward coupling 
 induced by transverse (diffraction) effects.
The result,
 while unremarkable, 
 does clarify the bounds on the validity
 of propagation models using a uni-directional approximation.

Since linear dispersion and finite nonlinear response times 
 will typically diminish any generation of a backward wave, 
 it is clear that any model which can be assumed uni-directional on 
 the basis of this paper will be more so in practice.
These results do not tell us whether
 the uni-directional approximation
 would be more or less robust for situations requiring vector fields
 (e.g. \cite{Maurer-JFBR-2007njp}), 
 or for nonlinear effects such as self-focusing \cite{Cumberbatch-1970imajmp}, 
 or nonlinear diffraction \cite{Boardman-MPS-2000oqe}
 but they at least establish a point of reference.

\section{Basic theory}
\label{S-basic}

Most optical pulse problems consider 
 a uniform and source free dielectric medium.
In such cases a good starting point 
 is the second order wave equation, 
 which results from the substitution of the $\nabla \times \vec{H}$
 Maxwell's equation into the $\nabla \times \vec{E}$ one
 in the source-free case
 (see e.g. \cite{Agrawal-NFO}).
Further, 
 assuming linearly polarized pulses, 
 we can use a scalar form.
Defining 
 $\nabla^2 = \partial_x^2 + \partial_y^2 + \partial_z^2$
 and $\partial_a \equiv \partial / \partial a$, 
 we can write the wave equation as
~
\begin{align}
  \left[
    \nabla^2
   -
    \frac{1}{c^2}
    \partial_t^2
  \right]
  E(t)
&=
  \frac{4\pi}
       {c^2}
  \partial_t^2
  P_T(E(t),t)
.
\label{eqn-basic-nabla2E}
\end{align}
Here I have suppressed the spatial coordinates for notational simplicity; 
 in fact we have $E(t) \equiv E(t,\vec{r})$ 
 and the total polarization $P(E(t),t) \equiv P_T(E(t,\vec{r}),t,\vec{r})$; 
 also $\vec{r} = (x,y,z)$.
Here we will consider only isotropic linear media, 
 which enables us to replace $P_T$ with a refractive index; 
 more complicated polarization behaviour is covered elsewhere
 \cite{Kinsler-2007-envel,Kinsler-2008-fchhg}.
Thus, 
 in the frequency domain, 
 we can write
~
\begin{align}
  \left[
    \nabla^2
   -
    \frac{n^2(\omega) \omega^2}{c^2}
  \right]
  E(\omega)
&=
  0
.
\label{eqn-basic-nabla2Ew}
\end{align}

However, 
 in most descriptions of pulse propagation we will want to chose a 
 specific propagation direction (e.g. along the $z$-axis), 
 and then denote the orthogonal components (i.e. along $x$ and $y$) 
 as transverse behaviour.
Thus, 
 splitting the $\nabla^2$ operator into its
 propagation ($z$) and transverse ($x, y$) parts, 
 we can rewrite the wave equation similarly:
~
\begin{align}
  \left[
    \partial_z^2 
   +
    K^2
  \right]
  E(\omega)
&=
 -
  \nabla_\perp^2
  E(\omega)
\label{eqn-bi-d2zE}
\\
  \left[
    \partial_z^2 
   +
    K^2
  \right]
  E(\omega)
&=
 +
  k_\perp^2
  E(\omega)
,
\label{eqn-bi-k2E}
\end{align}
where the total wavevector is given by $K^2 = n^2 \omega^2 / c^2$; 
 the transverse component is $k_\perp$.
If we want to describe diffraction, 
 then we can give the field some suitable beam profile $E(x,y)$, 
 an even simpler case is that of off-axis one dimensional propagation, 
 which requires merely a fixed value of $k_\perp$.

I now factorize the wave equation, 
 a process which, 
 while used in optics for some time \cite{Blow-W-1989jqe}
 has only recently been used to its full potential
 \cite{Ferrando-ZCBM-2005pre,Genty-KKD-2007oe,Kinsler-2007josab}.
Factorization takes its name from the fact that 
 the LHS of eqn.\eqref{eqn-bi-d2zE}
 is a simple difference of squares which might be factorized, 
 indeed this is what was done in a somewhat ad hoc fashion 
 by Blow and Wood in 1989 \cite{Blow-W-1989jqe}.
Since the factors are just $\partial_z \mp \imath K$, 
 we can see that each (by itself) would generate
 a forward directed wave equation, 
 and the other a backward one.
Without going into detail
 (although see the appendix), 
 a rigorous factorization procedure
 \cite{Ferrando-ZCBM-2005pre,Kinsler-2007-envel}
 allows us to define a pair of counter-propagating Greens functions, 
 and so divide the second order wave equation
 into a pair of coupled \emph{counter-propagating} first order ones.


Counter-propagating wave equations suggest counter propagating fields, 
 so I split the electric field up accordingly into forward ($E^+$) 
 and backward ($E^-$) parts, 
 with $E = E^+ + E^-$.
The coupled first order wave equations are
~
\begin{align}
    \partial_z
  E^\pm
&=
 \pm
    \imath
    K
  E^\pm
 \quad
 \mp
  \frac{\imath k_\perp^2}{2K}
  \left[
    E^+ + E^-
  \right]
.
\label{eqn-bi-dzE}
\end{align}
The RHS now falls into two parts, 
 which I term the underlying and residual parts \cite{Kinsler-2008-vpvg}.
First, 
 there is the $\imath K E^\pm$ term
 that, 
 by itself, 
 would describe plane-wave like propagation.
Second, 
 the remaining part (here proportional to $k_\perp^2$)
 which can be called ``residual'' terms.
These residual contributions, 
 here containing the transverse effects,
 account for the discrepancy between the true propagation
 and the underlying propagation.
Although here the residual component will be only a weak perturbation
 in e.g. the paraxial limit, 
 the theory presented here is valid for \emph{any} strength.
Although my preference would be to use a directional fields approach
 \cite{Kinsler-RN-2005pra} rather than the factorization
 one used here, 
 it is difficult to describe transverse effects satisfactorally.

Note that the work of Weston examines this kind of wave-splitting
 with more mathematical rigour
 (see e.g. \cite{Weston-1993jmp}), 
 although without consideration of residual terms, 
 and (at least initially) in the context of reflections and scattering.
This theory was based on that from the earlier work of 
 Beezley and Krueger \cite{Beezley-K-1985jmp} 
 who applied wave-splitting concepts to optics.

\subsection{Bi-directional (exact) case}
\label{S-bi}

The scalar second order wave equation
 given above in eqn.\eqref{eqn-basic-nabla2E}
 trivially provides a total wavector for any given 
 direction of propagation.
This is simply a sum of squares 
 of the parallel and transverse contributions
 $k_\parallel$ and $k_\perp$, 
 so that 
~
\begin{align}
  K^2
&=
  k_\parallel^2
 +
  k_\perp^2
.
\label{eqn-bi-K}
\end{align}

Note the reversed signs between the RHS terms 
 of eqn.\eqref{eqn-bi-dzE}, 
 the transverse part retards the propagation
 given by the underlying part; 
 the net forward-direction wavevector for some $k_\perp$
 is therefore less than the total wave vector --
 just as would be expected.

\subsection{Uni-directional approximation}
\label{S-uni}

Now I make the uni-directional assumption and 
 set $E^-=0$ in eqn.\eqref{eqn-bi-dzE},
 so that we get a wave equation with underlying ($\propto K$)
 and residual ($\propto k_\perp^2/2K$) components, 
 i.e.
~
\begin{align}
    \partial_z
  E^\pm
&=
 \pm
    \imath
    K
  E^\pm
 \quad
 \mp
  \frac{\imath k_\perp^2}{2K}
    E^+
.
\label{eqn-uni-dzE}
\end{align}
Note that the diffaction term here
 is identical to that obtained by applying the standard paraxial approximation 
 to propagation in a linear dispersive medium\footnote{
  I. M. Besieris, private communication}.

Alternatively, 
 since there is no $E^-$ field to complicate matters, 
 I might rewrite this wave equation
 using a new uni-directional wave vector $K_u$ to
define \emph{only} an underlying propagation,
 i.e. 
~
\begin{align}
    \partial_z
  E^\pm
&=
   \pm
    \imath
    K_u
  E^\pm
,
\label{eqn-uni-dzEK}
\end{align}
where $K_u$ is
~
\begin{align}
  K_u
&=
  K
 -
  \frac{k_\perp^2}{2K}
\quad
=
  K
  \left[
    1
   -
    \frac{1}{2}
    \frac{k_\perp^2}{K^2}
  \right]
.
\label{eqn-uni-Ku}
\end{align}
As expected, 
 $K_u$ is not equivalent to the true wave vector $K$; 
 indeed we expect it to be an approximation to $k_\parallel$,
 which specifies on-axis spatial variation of the exact propagation.


%
\section{Forward-backward coupling}
\label{S-coupling}

In the above, 
 we saw that the bi-directional and uni-directional models
 gave different propagation wave vectors.
However, 
 note that when $k_\perp^2/K^2 \ll 1$, 
 terms of order $(k_\perp/K)^4$ or higher 
 are negligible.
We can rearrange and then approximate eqn.\eqref{eqn-bi-K}
 in that limit so that 
~
\begin{align}
  k_\parallel^2
&=
   K^2
 -
  k_\perp^2
\\
  k_\parallel
&=
  K
  \left[
    1
   -
    \frac{k_\perp^2}{K^2}
  \right]^{1/2}
\\
&\simeq
  K
  \left[
    1
   -
    \frac{1}{2}
    \frac{k_\perp^2}{K^2}
  \right]
\quad
=
  K_u
.
\end{align}

Essentially what the condition $k_\perp^2/K^2 \ll 1$
 means is that propagation effects
 transverse to the chosen propagation direction must occur
 on a scale much larger than one wavelength, 
 or else a uni-directional approximation will fail.

It is important to note that the existence
 of significant forward-backward coupling
 does not always demand the presence or generation 
 of a freely propagating backward wave.
It is possible for the backward wave (i.e. $E^-$) to be dragged 
 along by the forward one, 
 as seen for several example in the directional fields formalism of 
 Kinsler et al. \cite{Kinsler-RN-2005pra}.
Nevertheless, 
 although it such a situation might give an answer
 correct to within a suitable scaling, 
 in such a case the uni-directional approximation is not 
 strictly valid.


%
\section{Conclusion}
\label{S-Conclude}

I have demonstrated one of the fundamental limits on the 
 widely used uni-directional propagation approximation.
This was done by a simple comparison of wave vectors 
 obtained from electromagnetic scalar wave equations
 allowing for all three spatial dimensions; 
 using both an exact (and hence bi-directional) model,
 and an approximate uni-directional model.
These results are done in the same style as, 
 and are intended to complement existing limits placed
 on nonlinear effects \cite{Kinsler-2007josab}; 
 they are not intended to startle the reader with their novelty.

I have shown that the condition
 $k_\perp/K^2 \ll 1$
 must hold for the uni-directional approximation to be true; 
 even when no backward field is initially present. 
Unsurprisingly this is comparable to the condition
 for the widely used ``paraxial'' limit.
Note, 
 however, 
 that the use of the paraxial approximation is rarely accompanied
 by a discussion of potential generation of backward propagating waves.

%

\acknowledgments

Thanks to I. M. Besieres for helpful comments.

%


%
\section*{Appendix: Factorizing}
\label{S-factorize}

Here is a quick derivation of the factorization process; 
 the $z$-derivative has been converted to $\imath k$, 
 $\beta^2 = n^2 \omega^2 /c^2$,
 and the unspecified residual term is denoted $Q$.
~
\begin{align}
  \left[
   -k^2 + \beta^2
  \right]
  E
&=
 -Q
\\
  E
&=
  \frac{1}{k^2 - \beta^2}
  Q
\qquad
=
  \frac{1}{\left(k-\beta\right)\left(k+\beta\right)}
\\
&=
  \frac{-1}{2\beta}
  \left[
    \frac{1}{k+\beta}
   -
    \frac{1}{k-\beta}
  \right]
  Q
.
\end{align}
Now $(k-\beta)^{-1}$ is a forward-like propagator for the field, 
 and $(k+\beta)^{-1}$ a backward-like propagator.
Hence write $E=E^++E^-$, 
 and split the two sides up
~
\begin{align}
  E^+
 +
  E^-
&=
  \frac{-1}{2\beta}
  \left[
    \frac{1}{k+\beta}
   -
    \frac{1}{k-\beta}
  \right]
  Q
\\
  E^\pm
&=
  \frac{\pm1}{2\beta}
    \frac{1}{k\mp\beta}
  Q
\\
  \left[
   k \mp \beta
  \right]
  E^\pm
&=
 \pm
  \frac{1}{2\beta}
    \frac{1}{k\mp\beta}
  Q
\\
  \imath
  k E^\pm
&= 
 \pm
  \imath
  \beta E^\pm
 \pm
  \frac{\imath}{2\beta}
  Q
,
\end{align}
and reverting to $z$ derivatives gives us the final form
~
\begin{align}
  \partial_z
  E^\pm
&= 
 \pm
  \imath
  \beta E^\pm
 \pm
  \frac{\imath}{2\beta}
  Q
.
\end{align}

\end{document}